\documentclass{vgtc}                          % final (conference style)
\graphicspath{{figures/}{pictures/}{images/}{./}} % where to search for the images

\usepackage{times}                     % we use Times as the main font
\usepackage{tabu}                      % only used for the table example
\usepackage{booktabs}                  % only used for the table example
\usepackage{lipsum}                    % used to generate placeholder text
\usepackage{mwe}                       % used to generate placeholder figures
\usepackage{soul}
\usepackage{graphicx} 
\usepackage{caption}
\usepackage{float}
\usepackage{afterpage}
\usepackage{accessibility}

\usepackage{mathptmx}                  % use matching math font

\onlineid{1074}

\vgtccategory{Poster}

\vgtcinsertpkg

\title{CvhSlicer 2.0: Immersive and Interactive Visualization of \\Chinese Visible Human Data in XR Environments}

\author{Yue Qiu$^{*}$, Yuqi Tong\thanks{Equal contributions.} , Yu Zhang, Qixuan Liu, Jialun Pei, Shi Qiu\thanks{Corresponding author: shiqiu@cse.cuhk.edu.hk}, Pheng-Ann Heng, Chi-Wing Fu\\ %
        \scriptsize Department of Computer Science and Engineering, The Chinese University of Hong Kong\\
        \scriptsize Institute of Medical Intelligence and XR, The Chinese University of Hong Kong}
\teaser{
  \centering
    \makebox[\textwidth]{%
        \includegraphics[width=0.93\linewidth, alt={The framework of CvhSlicer 2.0,  three main modules of data processing, 2D/3D visualization, and user interaction for XR-based immersive anatomical analysis.}]{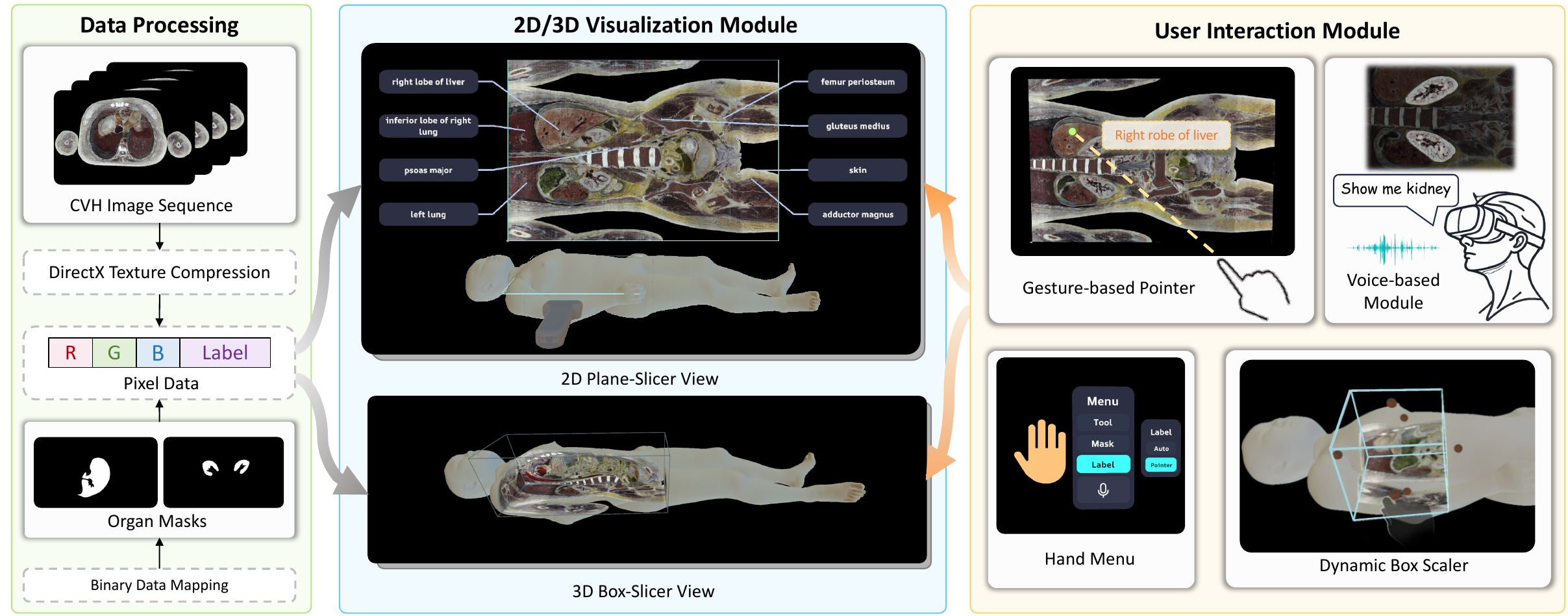}
    }
    \caption{The framework of the CvhSlicer 2.0 system. The system integrates data processing, 2D/3D visualization, and user interaction modules, introducing a novel pipeline for immersive and interactive anatomical data visualization.}
  \label{fig:framework}
}

\abstract{
    The study of human anatomy through advanced visualization techniques is crucial for medical research and education. In this work, we introduce CvhSlicer 2.0, an innovative XR system designed for immersive and interactive visualization of the Chinese Visible Human (CVH) dataset. Particularly, our proposed system operates entirely on a commercial XR headset, offering a range of visualization and interaction tools for dynamic 2D and 3D data exploration. By conducting comprehensive evaluations, our CvhSlicer 2.0 demonstrates strong capabilities in visualizing anatomical data, enhancing user engagement and improving educational effectiveness. A demo video is available at \url{https://youtu.be/CfR72S_0N-4}.

} % end of abstract


\begin{document}

\firstsection{Introduction}

\maketitle

Visualization of anatomical data is essential for modern medical research and education, providing critical insights into human anatomy and pathology. Among available resources, visible human data~\cite{ackerman1998visible,zhang2006chinese} stands out for its high-resolution, full-color cross-sectional images, enabling detailed exploration of complex anatomical structures. These realistic representations of human anatomy not only enhance understanding of tissue pathology and treatment outcomes but also support the creation of high-quality educational materials for medical training. To facilitate visualization of visible human data, tools such as \emph{CvhSlicer}~\cite{meng2011cvhslicer} have been developed. CvhSlicer achieves real-time navigation through cross-sectional anatomy by efficiently visualizing the large-scale Chinese Visible Human (CVH) dataset~\cite{zhang2006chinese} on standard PCs. However, its hardware reliance, limited interaction, and lack of immersion hinder usability, highlighting the need for advanced platforms.
While XR enables immersive and portable solutions, challenges persist in managing large datasets, real-time rendering, and resource optimization.

In this work, we introduce \textbf{CvhSlicer 2.0}, an XR-based system specifically designed for immersive and interactive visualization of high-resolution CVH data. Unlike the old CvhSlicer~\cite{meng2011cvhslicer}, our system operates entirely on an XR headset, providing seamless 2D/3D visualization and freehand interaction to enhance both user experience and system performance. Key features include customizable 2D and 3D slicers for real-time cross-sectional rendering, supported by trilinear interpolation for smooth and accurate image presentation, and a tailored shader to highlight regions of interest while preserving anatomical context. To further improve usability, CvhSlicer 2.0 integrates gesture-based tools, dynamic scalers for intuitive visualization controls, and voice commands for accessible interaction. Leveraging advanced visualization, optimized rendering, and multimodal interaction, our system shows great potential to transform medical education, surgical training, and anatomical research.

\section{Implementation}

\subsection{Data Processing}
The CVH dataset~\cite{zhang2006chinese} includes 3,640 slices of a Chinese female cadaver, 0.25 mm spaced in the head-neck region and 0.5 mm in the rest of the body, with each slice as a high-resolution image of $3,872 \times 2,048$. To enhance computational efficiency, we implement several data processing steps to compress the dataset. First, we downsample the images, resulting in a lower resolution for each slice. Next, we employ binary data mapping to extract organ regions as masks, generating pixel-level annotations. Finally, we apply the DXT algorithm \cite{renambot2007real} to further compress the image blocks, benefiting from its minimal quality loss and rapid decompression. These steps reduce the overall dataset size from 35.7 GB to 3.0 GB, making it suitable for storage and loading on mobile XR devices. In practice, the data loading time on the Meta Quest 3 is $\sim$10 seconds.
\begin{figure}
    \centering
    \includegraphics[width=0.94\linewidth, alt={The evolution of CvhSlicer, showcasing an old PC-based system and a recent XR-based system for immersive and interactive visualization.}]{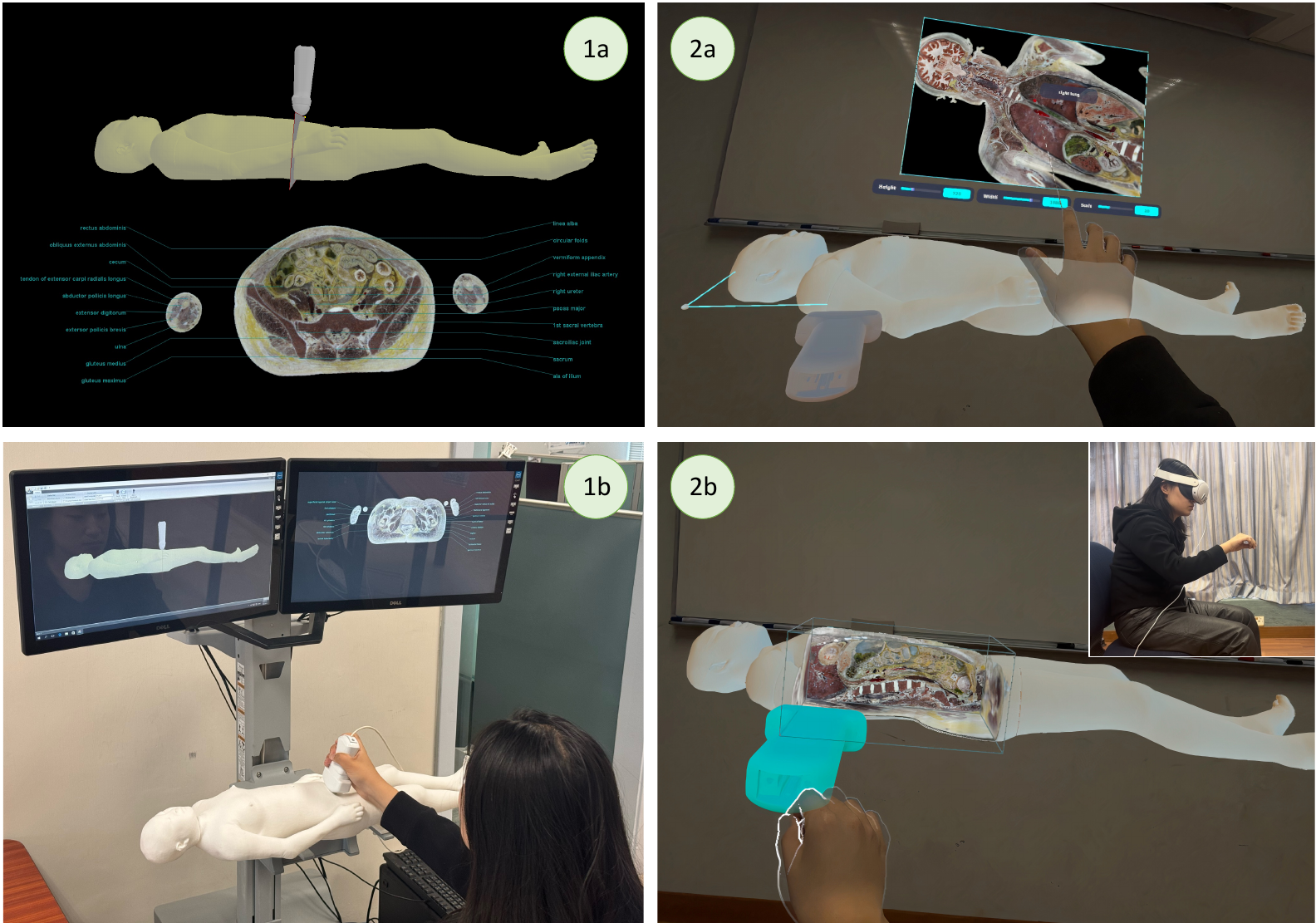}
    \caption{(1a) The traditional CvhSlicer~\cite{meng2011cvhslicer} interface for PC-based interaction. (1b) The hardware setup and user operation of the CvhSlicer system. (2a) The 2D plane-slicer of our CvhSlicer 2.0 in XR mode. (2b) The 3D box-slicer of our CvhSlicer 2.0 in XR mode.}
    \label{fig:enter-label}
\end{figure}
\subsection{2D/3D Visualization Module}
We visualize the CVH dataset in both 2D and 3D views. For 2D visualization, we implement a virtual transducer that allows users to position a plane-slicer within the 3D replica of the CVH body. As the plane-slicer intersects the volume data, we calculate and render the relevant information onto a virtual screen for visualization. Specifically, we compute the RGB values for each pixel using the CPU of the XR device and apply trilinear interpolation to sample the voxelized data, ensuring a smooth rendering. To maintain high frame rates during the movement of the slicer, we develop a dynamic resolution rendering technique that enables rapid rendering of frames at a lower resolution when necessary.

For 3D visualization, we employ a 3D box-slicer to probe the volumetric dataset. This method dynamically renders cross-sectional views of internal anatomical structures along the inner surfaces of the box, offering enhanced geometric insights into the anatomical data. We implement a customized shader to render sections of the human surface mesh intersected by the box-slicer, creating a realistic ``cutting” effect. Users can adjust the position, size, and orientation of the box to intuitively explore high-resolution structures while maintaining spatial context.

\subsection{User Interaction Module}

To intuitively learn the information of anatomical structures, we design two complementary label identification methods. The first method provides a comprehensive overview of major anatomical structures in each 2D cross-sectional slice, helping users easily identify key structures and gain a holistic understanding of the anatomical data. Specifically, we analyze the pixel count of each anatomical category in the slice to identify key structures, calculate their geometric centroids, and display category labels around the edges of the rendered slice. This approach ensures that users can quickly recognize and contextualize the primary anatomical features within each slice. For more detailed and localized exploration, we implement a gesture-based pointing method, which detects the user’s index finger position and projects a ray from it. By calculating the intersection of this ray with the rendered 2D-plane slicer, the system identifies the anatomical structure at the intersection point. Coupled with a gesture-controlled hand menu, we allow users to dynamically query and explore specific structures in real time, offering a hands-on and interactive experience.

To enhance interactivity, we integrate a voice control module using the Meta Voice SDK. It transcribes voice inputs into text commands, enabling dynamic visualization of anatomical structures through natural language. This voice-based interaction serves as a complementary mechanism to the gesture-based method, offering an alternative for precise structure identification. By combining gesture and voice controls, our multimodal interaction approach significantly enhances the usability of CvhSlicer 2.0. 

\section{Evaluation}
We conduct a pilot study with 10 participants ($n=10$), comprising 3 females and 7 males, all graduate students (average age $=27.3$) from engineering and medical schools. For each tested system, we first introduce its main operations and functionalities to the participants, and then ask them to freely interact with the visualized CVH data for $\sim$10 minutes. The study measures the System Usability Scale (SUS)~\cite{brooke1996sus} for both CvhSlicer and CvhSlicer 2.0 and collects qualitative feedback. The SUS results demonstrate a significant improvement in usability with \textbf{CvhSlicer 2.0}, which achieves an average score of \textbf{91.0 (SD = 6.14)}, compared to \emph{70.0 (SD = 22.89)} for the original \emph{CvhSlicer}. The higher score and notably lower standard deviation indicate that CvhSlicer 2.0 offers a more user-friendly experience and consistent performance.

Participants also provide qualitative feedback on novel system features. The visualization of anatomical data in CvhSlicer 2.0 is praised for its clarity and detail, with most users highlighting its effectiveness in displaying fine anatomical structures. Gesture-based tools are praised for accuracy and ease in identifying structures, while voice commands enhance convenience as an alternative to manual controls. Our immersive approach significantly improves spatial understanding, surpassing traditional systems.

\section{Conclusion and future work}
In this work, we present CvhSlicer 2.0, an XR-based system designed for immersive and interactive visualization of human anatomy. By integrating gesture and voice inputs, the system enables intuitive medical data navigation while addressing several limitations of existing methods. Key improvements include enhanced spatial visualization of anatomical data, intuitive freehand interaction, increased portability, and more precise engagement with anatomical structures. However, our collected user feedback highlights the need to improve the accuracy of voice commands, particularly for users with diverse accents. Future work will focus on refining this feature and exploring advanced data analysis and visualization techniques to expand the system's applicability, including its use in collaborative medical education scenarios.

\acknowledgments{This work was supported by the Research Grants Council of the Hong Kong Special Administrative Region, China (Project No.: T45-401/22-N).}

\bibliographystyle{abbrv}

\bibliography{ref}
\end{document}